\documentclass{aastex}
\usepackage{amsmath}
\usepackage{amssymb}
\usepackage[numberedappendix]{emulateapj5}
\usepackage{epsfig}

\makeatletter

\newenvironment{inlinefigure}{%
\def\@captype{figure}%
\noindent\begin{minipage}{0.999\linewidth}\begin{center}}
{\end{center}\end{minipage}\smallskip}
\makeatother

\newcommand{\kms}{\,{\rm km\,s^{-1}}}
\newcommand{\au}{\,{\rm AU}}
\renewcommand{\day}{\,{\rm d}}
\newcommand{\yr}{\,{\rm yr}}

\newcommand{\gyr}{\,{\rm Gyr}}

\newcommand{\pc}{\,{\rm pc}}
\newcommand{\kpc}{\,{\rm kpc}}

\newcommand{\ergs}{\,{\rm ergs\,s^{-1}}}
\newcommand{\s}{\,{\rm s}}

\newcommand{\msun}{\,M_\odot}
\newcommand{\rsun}{\,R_\odot}
\newcommand{\lx}{L_{\rm X}}
\newcommand{\mdot}{\,M_\odot\,{\rm yr}^{-1}}

\newcommand{\ace}{\eta_{\rm CE}}
\newcommand{\bbar}{\bar{B}}

\begin{document}

\shorttitle{A NEW CLASS OF HIGH-MASS X-RAY BINARIES}                                    
\shortauthors{PFAHL, RAPPAPORT, PODSIADLOWSKI, \& SPRUIT}

%%%%%%%%%%%%%%%%%%%%%%%%%%%%%%%%%%%%%%%%%%%%%%%%%%%%%%%%%%%%%%%%%%%

\submitted{Submitted to ApJ}

\title{A NEW CLASS OF HIGH-MASS X-RAY BINARIES: \\
IMPLICATIONS FOR CORE COLLAPSE AND NEUTRON-STAR RECOIL}

\author{Eric Pfahl and Saul Rappaport}
\affil{Department of Physics, Massachusetts Institute of Technology, Cambridge,
MA, 02139; \\ pfahl@space.mit.edu, sar@mit.edu}

\and

\author{Philipp Podsiadlowski}
\affil{Nuclear and Astrophysics Laboratory, Oxford University, Oxford, OX1 
3RH, England, UK; \\
podsi@astro.ox.ac.uk}

\and

\author{Hendrik Spruit}
\affil{Max-Planck-Institut f\"ur Astrophysik, Postfach 1317, D-85741,
Garching, Germany; \\
henk@mpa-garching.mpg.de}

%%%%%%%%%%%%%%%%%%%%%%%%%%%%%%%%%%%%%%%%%%%%%%%%%%%%%%%%%%%%%%%%%%%

\begin{abstract}

We investigate an interesting new class of high-mass X-ray binaries (HMXBs) with 
long orbital periods ($P_{\rm orb} > 30\day$) and low eccentricities ($e \la 0.2$).  
The orbital parameters suggest that the neutron stars in these systems did not receive 
a large impulse, or ``kick,'' at the time of formation.  After considering the statistical
significance of these new binaries, we develop a self-consistent phenomenological picture 
wherein the neutron stars born in the observed wide HMXBs receive only a small kick 
($\la 50 \kms$), while neutron stars born in isolation, in the majority of low-mass X-ray 
binaries, or in many of the well-known HMXBs with $P_{\rm orb} \la 30\day$ receive the 
conventional large kicks, with a mean speed of $\sim 300 \kms$.  Assuming that this basic 
scenario is correct, we discuss a physical process that lends support to our hypothesis, 
whereby the magnitude of the natal kick to a neutron star born in a binary system  depends 
on the rotation rate of its immediate progenitor following mass transfer --- the core of the 
initially more massive star in the binary.  Specifically, the model predicts that rapidly rotating 
pre-collapse cores produce NSs with relatively small kicks, and vice versa for slowly
rotating cores.  If the envelope of the NS progenitor is removed before it has become deeply convective,
then the exposed core is likely to be a rapid rotator.  However, if the progenitor becomes
highly evolved prior to mass transfer, then a strong magnetic torque, generated by differential
rotation between the core and the convective envelope, may cause the core to spin down to 
the very slow rotation rate of the envelope.  Our model, if basically correct, has important 
implications for the dynamics of stellar core collapse, the retention of neutron stars in globular 
clusters, and the formation of double neutron star systems in the Galaxy.

\end{abstract}

%%%%%%%%%%%%%%%%%%%%%%%%%%%%%%%%%%%%%%%%%%%%%%%%%%%%%%%%%%%%%%%%%%%

\keywords{stars: neutron --- supernovae: general --- X-rays: stars}

%%%%%%%%%%%%%%%%%%%%%%%%%%%%%%%%%%%%%%%%%%%%%%%%%%%%%%%%%%%%%%%%%%%

\section{INTRODUCTION}

It has become fashionable in recent years to suppose that the majority of 
neutron stars (NSs) are born with speeds in excess of $\sim 100-200 \kms$, presumably
as a result of some asymmetry in the core collapse or the subsequent supernova (SN)
explosion of the NS progenitor.  The strongest support for this notion comes from 
the high speeds inferred for the $\sim 100$ Galactic pulsars with well-measured 
interferometric proper motions (Harrison, Lyne, \& Anderson 1993).  
Mean speeds for these pulsars of $\ga 300 \kms$ have been estimated by a number of 
authors \citep[e.g.,][]{lyne94,hansen97,cordes98,arzoumanian01}.  Various classes of binary 
systems containing NSs also show strong evidence for substantial natal ``kick'' velocities, 
based upon their present orbital parameters, their systemic speeds, and/or their height
above the Galactic plane \citep[e.g.,][]{brandt95,verbunt95,johnston96,vdh00}.  

Very large uncertainties, both observational and theoretical, still pervade studies 
of the underlying distribution in NS natal kick speeds.  Complications include the fairly 
small sample of pulsars with proper-motion measurements, questionable dispersion-measure 
distances, serious observational selection effects, and uncertainties regarding the 
formation of NSs and their dynamical evolution in the Galaxy.  Fortunately, the sample
of pulsars with reliable proper motions is growing \citep[see][]{mcgary01}, as is the 
number of pulsars with accurate parallax distances \citep[e.g.,][]{toscano99,brisken00}.     

The most popular models for NS kicks involve a momentum impulse delivered around
the time of the core collapse that produced the NS.  Mechanisms in this class 
include purely hydrodynamical processes, as well as primarily 
neutrino-driven kicks (see Lai 2000 for a review).  In either case, some process must 
be responsible for breaking spherical symmetry during core collapse, such as a combination of 
Rayleigh-Taylor instabilities and neutrino-induced convection \citep[e.g.,][]{janka94,fryer00}.  
A fundamentally different mechanism for producing significant NS velocities was
proposed by \citet{harrison75}, whereby the NS is accelerated {\em after} 
the core-collapse event as a result of asymmetric electromagnetic (EM) dipole 
radiation --- the so-called EM ``rocket'' effect.  This process is 
distinctly non-impulsive.

A theoretical determination of the emergent velocity distribution associated
with each kick mechanism is extremely difficult and would require an ensemble of 
very detailed three-dimensional hydrodynamical simulations (with the exception of 
the EM ``rocket'' mechanism).  Furthermore, it is unlikely that a single process 
accounts for the full range of NS velocities.  For instance, it is plausible that the
dominant kick mechanism and the magnitude of the kick depend at least somewhat
on the evolutionary history of the NS progenitor.  In this paper, we explore
a possible linkage between the kick magnitude and the evolution of the NS progenitor
in a binary system.  This work was inspired by a new observed sub-class of high-mass 
X-ray binaries (HMXBs).

Previously, significant eccentricities of $e \sim 0.3-0.5$ seemed to be the general rule
among HMXBs with $P_{\rm orb} \sim 20-100\day$ \citep[see][]{bildsten97}, which is presumably
the result of a substantial NS kick \citep[e.g.,][]{brandt95,verbunt95,vdh00}.  By contrast, 
the new class of HMXBs are clearly distinguished by their low eccentricities of $e \la 0.2$ and long 
orbital periods of $P_{\rm orb} \sim 30 - 250\day$, which indicate that tidal circularization
should not have played a significant role if the massive stellar component is not very 
evolved.   Eccentricities of this magnitude are roughly consistent with the dynamical effect 
of mass loss alone in the SN explosion, although relatively small kick speeds of 
$\la 50 \kms$ cannot be ruled out on statistical grounds.  

There are currently six candidates for the new class of HMXBs. This is a substantial
number, given the difficulties associated with detecting these binaries and measuring
their orbits, and the fact that there are only $\sim 20$ HMXBs with measured orbital 
parameters.  We suggest that the observed wide, nearly circular HMXBs are representative
of a much larger intrinsic population, and that the NSs in these systems received only a
small kick ($\la 50 \kms$).  We further speculate that the magnitude of the kick is correlated 
with the evolutionary history of the binary system, before the formation of the NS.  
Specifically, we propose that the kick speed depends on the rotation rate of the core of the 
NS progenitor following a phase of mass transfer, wherein the hydrogen-rich envelope of the 
star is removed.  The sense is that slowly rotating cores produce NSs with the conventional
large kicks, while the collapse of rapidly rotating cores are accompanied by relatively small
natal kicks.  If our basic picture is correct, there may be important implications for magnetic 
field evolution and core collapse in massive stars, the retention of NSs in globular clusters, 
and the birthrate of double NS binaries in the Galaxy.  

In \S~\ref{sec:newbin}, we discuss the observed characteristics of the new 
class of wide, low-eccentricity HMXBs.  A brief theoretical overview of the formation, evolution, 
and population synthesis of massive binaries is given in \S~\ref{sec:popsyn}.
Using a combination of theoretical and observational arguments, we claim in \S~\ref{sec:sig}
that mean kick speeds of $\ga 200-300\kms$ are not consistent with the numbers
and properties of the members of the new observed class of HMXBs, and that considerably smaller 
kicks of $\la 50 \kms$ are probably required.  We develop in \S~\ref{sec:mod} 
a phenomenological picture that accounts for the new HMXBs, and which is consistent with what 
is known about the Galactic NS populations.  We lend some credence to this basic picture by 
suggesting a plausible physical scenario in \S~\ref{sec:phys} that naturally relates the 
rotation rate of the pre-collapse core (NS progenitor) and the evolutionary history of its
host binary system.  Finally, we investigate in \S~\ref{sec:imp} a number of further implications 
of the new class of HMXBs and our associated model.  Our main points are summarized in 
\S~\ref{sec:con}.     

%%%%%%%%%%%%%%%%%%%%%%%%%%%%%%%%%%%%%%%%%%%%%%%%%%%%%%%%%%%%%%%%%%%

\section{A NEW CLASS OF HIGH-MASS X-RAY BINARIES}\label{sec:newbin}

\begin{figure*}[t]
\centerline{\epsfig{file=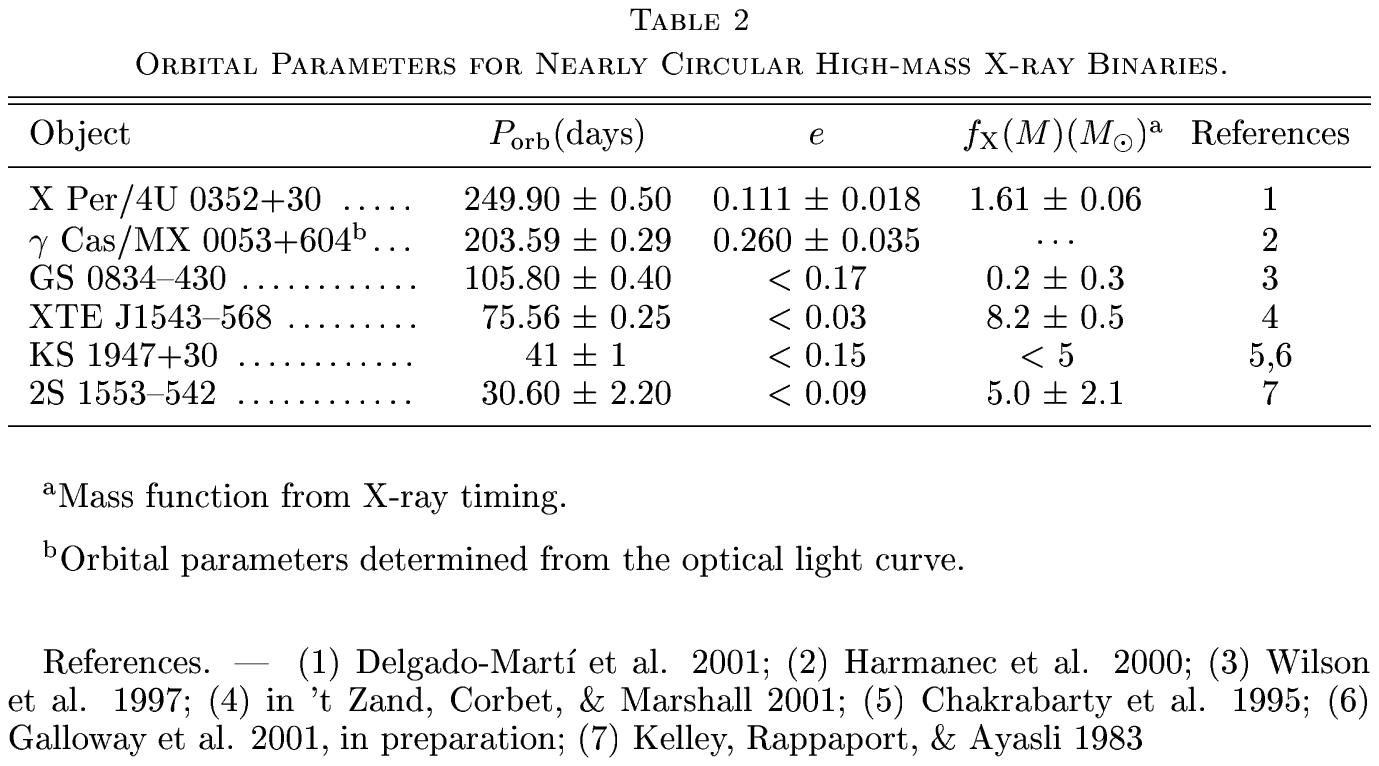,width=0.7\linewidth}}
\end{figure*}

A HMXB consists of a NS, which often appears as an X-ray pulsar, and a massive stellar
companion.  Of the $\sim 130$ known HMXBs \citep[see][]{liu00}, 
$\sim 20$ have reasonably well-measured orbital elements (see Bildsten et al. 1997 for a 
somewhat dated list).  In all but one case ($\gamma$ Cas; Harmanec et al. 2000), 
the parameters were determined from the timing of the X-ray pulsar.    

Two classes of HMXBs, distinguished by their orbital parameters, are apparent in 
Table 3 of Bildsten et al. (1997): (i) systems with $P_{\rm orb} \la 10\day$ and 
$e \la 0.1$, and (ii) moderately wide, eccentric binaries with 
$P_{\rm orb} \sim 20 - 100\day$ and $e \sim 0.3 - 0.5$.  
A new third class of HMXBs has recently emerged.  These systems are distinguished 
from the well-known HMXBs by their wide orbits (all have $P_{\rm orb} > 30\day$)
and fairly low eccentricities ($e \la 0.2$).  Table 1 lists the names and orbital 
parameters of these interesting binaries, and below we give a brief synopsis 
of relevant observational information for each system.  For two of the binary X-ray pulsars
discussed below (XTE J1543--569 and 2S 1553--542), the optical counterpart has 
not been identified.  In these cases, we should remain open to the possibility that 
the companion may have evolved beyond the main sequence and is filling a significant 
fraction of its Roche lobe, thus explaining the low eccentricities.  

%%%%%%%%%%%%%%%%%%%%%%%%%%%%%%%%%%%%%%%%%%%%%%%%%%%%%%%%%%%%%%%%%%%

\bigskip

\subsection{Observational Information for Each Binary}\label{sec:ind} 

\subsubsection{X Per/4U 0352+309}

The X-ray counterpart to the classical Be (or possibly Oe) star X Per, 4U 0352+309, 
exhibits pulsations with a period of $\sim 837 \s$.  Variations in the pulse period 
strongly suggest that the X-ray source is an accreting NS.  The X-ray pulsar was 
observed by \citet{delgado01} over an interval of nearly 600 days with the 
{\em Rossi X-ray Timing Explorer} ({\em RXTE}).  These observations have revealed the 
orbital period of the system, $P_{\rm orb}=250.3 \pm 0.6\day$, and the orbital 
eccentricity, $e=0.111 \pm 0.018$.

Estimates of the distance to X Per range from $700 \pm 300 \pc$ to $1.3 \pm 0.4 \pc$
(see Delgado-Mart\'\i~et al. 2001 and references therein).  It is then especially 
interesting to note that X Per lies at a Galactic latitude of approximately $-17 \degr$.  
For an assumed distance of $\sim 1 \kpc$, this latitude places X Per at a height of 
$\sim 300 \pc$ above the Galactic plane, which is much larger than the scaleheight of 
early-type stars in the Galactic disk.  This large height may be explained by the systemic 
impulse received due to the mass loss and kick associated with the formation of the NS.  
However, the magnitude of the kick would have to be quite large, and the near circularity 
of the orbit would make the X Per system a very unlikely object.  A far simpler and 
more reasonable hypothesis is that the binary was, in fact, born in an OB association 
within the Gould Belt \citep[e.g.,][]{torra00}, a disk-like structure with a radial 
extent of $\ga 500 \pc$, inclined by $\sim 20 \degr$ to the Galactic plane.  It is 
thought that the associations comprising the Gould Belt account for roughly $60\%$ of 
the O and B stars within $\sim 500 \pc$ from the Sun.
                  
\subsubsection{$\gamma$ Cas/MX 0053+604}

It has long been suspected that $\gamma$ Cas, the first-known Be star \citep{secchi67}, 
is a member of a binary system; however, the orbit has
defied detection at both X-ray and optical wavelengths until very recently.
\citet{harmanec00} have measured the orbit of the $\gamma$ Cas system using optical 
spectroscopy.  Periodic shifts in H$\alpha$ and He I line features were attributed to 
the orbital motion of the Be star.  The H$\alpha$ measurements yielded the orbital 
parameters $P_{\rm orb} = 203.59 \pm 0.29\day$ and $e = 0.26 \pm 0.035$.  The optical
mass function implies that the unseen companion has a mass of $\sim 1 \msun$, consistent
with a massive white dwarf or a NS.  

There is still debate regarding the nature of the X-ray counterpart to $\gamma$ Cas,
MX 0053+60.  If the companion is indeed a compact object, it is not clear from the X-ray
emission whether it is a NS or a white dwarf (e.g., no pulsations have been detected).  
However, if the system presently contains a white dwarf, we would expect the orbit to be 
circular as a result of an earlier episode of mass transfer.  On these grounds, the NS
hypothesis is compelling, since the SN that accompanied the formation of the NS could have 
easily perturbed the orbit to yield the observed eccentricity.  Smith, Robinson, \& Corbet 
(1998; see also Robinson \& Smith 2000) argue against the hypothesis that the X-rays emanate 
from a compact object and favor a model where the X-ray emission is the result of 
magnetic activity on the stellar surface.  Further observations are required to determine 
the origin of the X-rays and the nature of the companion to $\gamma$ Cas.

\bigskip

\subsubsection{GS 0834--430}

\citet{wilson97} analyzed the data from seven outbursts of the transient X-ray pulsar 
GS 0834--430, observed with the BATSE instrument on board the {\em Compton Gamma Ray
Observatory} ({\em CGRO}).  Timing analysis of the 12.3 s X-ray pulsar revealed an orbital period
of $P_{\rm orb} = 105.8 \pm 0.4\day$, but did not place very tight constraints on the 
eccentricity.  According to \citet{wilson97}, a likely value for the eccentricity is
$e \la 0.2$; larger values are permitted, but require a very small binary inclination.
From the spin-up behavior of the X-ray pulsar, the estimated distance of the binary is 
$\ga 4.5 \kpc$.  The optical counterpart to GS 0834--430 has been identified as a Be 
star by \citet{israel00}

\subsubsection{XTE J1543--569}

After a year-long monitoring campaign with {\em RXTE}, \citet{zand01} have determined 
the orbital parameters of the X-ray pulsar XTE J1543--569.  The system has an orbital 
period of $P_{\rm orb} = 75.56 \pm 0.25\day$ and an eccentricity of $e < 0.03$ at the 
$2\,\sigma$ level.  This eccentricity is surprisingly small if the massive
companion to the NS is near the main sequence and thus is greatly underfilling its 
Roche lobe, even if we assume that the NS did not receive a kick.  However, an optical 
counterpart has yet to be discovered, although the orbital and pulse periods place 
XTE J1543--569 amongst the confirmed Be/X-ray transients in the ``Corbet'' diagram 
(Corbet 1986; see also Bildsten et al. 1997).  

It is interesting to note that the present eccentricity of XTE J1543--569 is not
likely to be consistent with a vanishing NS kick.  If we consider only mass loss
in the SN explosion, the induced eccentricity for an initially circular orbit is
$e = \Delta M/(M_b - \Delta M)$, where $\Delta M$ is the mass lost and $M_b$ is the
pre-SN mass of the binary \citep[e.g.,][]{blaauw61,dewey87}.  An eccentricity of $\sim 0.03$ 
is obtained if $\Delta M = 0.6\msun$, for a pre-collapse core mass of $2\msun$, and 
$M_b = 20\msun$, a somewhat high but not very unlikely mass.  However, more typical values 
of $\Delta M$ and $M_b$ are $1.6\msun$ and $15\msun$, respectively, which yield $e \sim 0.12$.  
In this case, a kick is required to ``correct'' the eccentricity to produce the smaller 
observed value, but the magnitude and direction of the kick must be somewhat finely tuned.  
We suggest that possibly either the stellar companion to the X-ray source XTE J1543--569 is very 
massive or that the companion has evolved well beyond the main sequence and is filling a sizable 
fraction of its Roche lobe, so that tidal circularization accounts for the low eccentricity.

\subsubsection{KS 1947+30}

The transient X-ray source KS 1947+30 was first detected by the Kvant instrument 
on board the {\em Mir} space station \citep{borozdin90}.  Later, in 1994, 
18.7\,s X-ray pulsations were detected by BATSE during an outburst that lasted 33\,d 
(Chakrabarty 1995 and references therein).  However, the $\sim 10\,{\rm deg}^2$ position 
resolution of BATSE was not sufficient to identify the X-ray pulsar as the Kvant source, 
and the pulsar was given the designation GRO J1948+32.  Modulation of the pulse frequency during the 
33\,d outburst was suggestive of a binary orbit, but with less than one full orbital 
cycle of coverage.  Preliminary estimates placed the orbital parameters in the ranges  
$35 \day < P_{\rm orb} < 70 \day$ and $e < 0.25$

A recent outburst has allowed KS 1947+30 to be ``rediscovered'' by the All Sky Monitor (ASM)
on board {\em RXTE} (Galloway et al. 2001, in preparation).  It was quickly realized that GRO J1948+32 
and the old Kvant source are the same, and so the earlier designation, KS 1947+30,
has been adopted.  The X-ray pulsar has now been timed for $\sim 5$ orbits, and a precise
orbital solution has been determined, with $P_{\rm orb} = 41.12 \pm 0.65\day$ and 
$e = 0.12 \pm 0.02$.  Furthermore, the accurate position has led to the identification 
of an optical counterpart, probably an O/Be star \citep{negueruela00}.               

\subsubsection{2S 1553--542}

The transient X-ray pulsar 2S 1553--542 was first detected with the {\em SAS 3} satellite 
during the only known outburst of the source in 1975 \citep[see][]{apparao78}. 
\citet{kelley83} analyzed data that spanned 20\,d of the outburst and discovered regular
variations in the 9.27\,s pulse period that they attributed to a binary orbit.  They determined 
that the system has an orbital period of $P_{\rm orb} = 30.6 \pm 2.2\day$ and an 
eccentricity of $e < 0.09$.  The orbital parameters were not well constrained
because the observations did not cover a full orbital cycle.   
Although no optical counterpart to 2S 1553--542 has been identified, the highly transient
nature of the source is suggestive of an unevolved Be star companion
(see Kelley, Rappaport, \& Ayasli 1983 and references therein).

%%%%%%%%%%%%%%%%%%%%%%%%%%%%%%%%%%%%%%%%%%%%%%%%%%%%%%%%%%%%%%%%%%%

\section{AN OVERVIEW OF MASSIVE BINARY \\ POPULATION SYNTHESIS}\label{sec:popsyn}

From a theoretical point of view, the formation of a NS in a binary system involves 
three distinct evolutionary steps: (1) the formation of a primordial 
binary, where the initially more massive component (the primary) has a mass
$\ga 8 \msun$, (2) a phase of mass transfer from the primary to the 
secondary\footnote{Hereafter, the term ``secondary'' will refer to the initially
less massive star, whether or not the secondary has become the more massive component
of the binary system as a result of mass accretion.}
(the initially less massive component), which may be dynamically unstable, 
and (3) the subsequent SN explosion of the primary's hydrogen-exhausted core 
and the formation of the NS, where the disruptive influence of the SN may unbind 
the binary system.  None of these steps is especially well understood, and so we 
encapsulate our lack of detailed knowledge in the form of a set of free parameters, 
some of which have values that are constrained by observations.  In this paper, 
we present our results for one standard-model set of parameters.  We have varied 
of number of the free parameters in our study and found that our main results and
conclusions are unchanged.  We now give a very brief overview of the
elements of our Monte Carlo binary population synthesis code.  A far more detailed 
discussion and an extensive set of references is provided in 
Pfahl, Rappaport, \& Podsiadlowski (2001; hereafter, PRP).  

The primary mass, $M_1$, is chosen from a power-law distribution, $p(M_1) \propto M_1^{-x}$,
which is appropriate for massive stars \citep[see][]{miller79,scalo86,kroupa93}.
For our standard model, we choose $x=2.5$, where $x=2.35$ corresponds to a 
Salpeter IMF \citep{salpeter55}.  An isolated star with solar metallicity will produce
a NS if its mass is between $\sim 8$ and $30 \msun$, where the upper limit is quite 
uncertain, but has little impact on our results.  The secondary mass, $M_2$, is assumed to 
be correlated with the primary mass according to a distribution in mass ratios, 
$p(q) \propto q^y$, where $q \equiv M_2/M_1 < 1$.  We adopt a flat distribution in 
mass ratios ($y=0$) for our standard model.

We assume that the massive primordial binary is circular (see the remarks in PRP)
and draw the orbital separation, $a$, from a distribution that is uniform in the
logarithm of $a$ \citep[e.g.,][]{abt78}.  The minimum value of $a$ is determined from the
constraint that neither star overflows its Roche lobe on the main sequence.  The upper
limit should be large, but is otherwise arbitrary, and its value does not significantly 
affect our results.  In practice, we choose a maximum separation of $10^3 \au$. 

If the orbit of the primordial binary is sufficiently compact, the primary will grow
to fill its Roche lobe, where the volume-equivalent radius of the Roche lobe about the 
primary is approximated by the formula due to \citet{eggleton83}:
\begin{equation}\label{eq:roche}
\frac{R_{\rm L1}}{a} \equiv r_{\rm L1} = 
\frac{ 0.49 }{ 0.6+q^{2/3}\ln(1+q^{-1/3}) } ~.
\end{equation}   
The evolutionary state of the primary when it fills its Roche lobe, in conjunction with 
the mass ratio of the components, is a good indicator of the physical character of the 
subsequent mass transfer and binary stellar evolution.  It is common practice to distinguish 
among three evolutionary phases of the primary at the onset of mass transfer, following  
Kippenhahn \& Weigert 1966 (see also Lauterborn 1970; Podsiadlowski, Joss,
\& Hsu 1992).  Case A evolution corresponds to core hydrogen burning, Case B refers 
to the shell hydrogen-burning phase, but prior to central helium ignition, and case C 
evolution begins after helium has been depleted in the core.  A large fraction of binaries 
will be sufficiently wide that the primary and secondary evolve essentially as isolated stars
prior to the first SN.  We refer to such detached configurations as case D.
Cases A, B, C, and D comprise roughly $5\%$, $25\%$, $25\%$, and $45\%$, respectively,
of the primordial binary population with our standard-model parameters.

It is particularly important to distinguish between mass transfer that is dynamically
{\em stable} (proceeding on the nuclear or thermal timescale of the primary) and mass
transfer that is dynamically {\em unstable} (proceeding on the dynamical timescale of the
primary).  Cases B and C are naturally divided into an {\em early} case (B$_e$ 
or C$_e$), where the envelope of the primary is mostly radiative, and a {\em late} 
case (B$_l$ or C$_l$), where the primary has a deep convective envelope.
We assume that cases B$_e$ and C$_e$ mass transfer are stable if the 
mass ratio ($q=M_2/M_1$) is greater than some critical value, $q_{\rm crit}$, which we take 
to be 0.5 in our standard model.  Cases B$_l$ and C$_l$ mass transfer are assumed 
to be dynamically unstable, regardless of the mass ratio.  The reader is directed
to PRP for further details.  

During stable mass transfer, some fraction, $\beta$, of the material lost by the 
primary through the inner Lagrange point (L1) is accreted by the secondary.  The 
excess material escapes the system with specific angular momentum $\alpha (GM_ba)^{1/2}$,
where $\alpha$ is a dimensionless parameter and $M_b = M_1 + M_2$; both $M_b$ and $a$ take their
instantaneous values during mass transfer.  A reasonable analytic description of the 
orbital evolution is obtained when $\alpha$ and $\beta$ ($>0$) are held fixed \citep[see][]{podsi92}:
\begin{equation}\label{eq:albe}
\frac{a'}{a} = 
\frac{M_b'}{M_b} \left( \frac{M_1'}{M_1} \right)^{C_1}
\left( \frac{M_2'}{M_2} \right)^{C_2} ~,
\end{equation}
where
\begin{eqnarray}
C_1 & \equiv & 2\alpha(1-\beta)-2 \nonumber \\ 
C_2 & \equiv & -2\alpha \left(\frac{1}{\beta}-1\right)-2 ~.
\end{eqnarray}
Primes on the masses and semimajor axis indicate the values after some amount of mass has 
been transferred.  For our standard model, we somewhat arbitrarily let $\beta = 0.75$, and 
we take $\alpha = 1.5$, a value characteristic of mass loss through the L2 point.

During stable mass transfer, the secondary will respond in one of two ways.
If accretion occurs while the secondary is still on the main sequence, the secondary will
generally be ``rejuvenated.''  That is, the evolutionary clock of the secondary will be 
reset (though not precisely to the zero-age main sequence), and its subsequent evolution will 
be very similar to the evolution of an isolated star with a larger mass (Hellings 1983; 
Podsiadlowski, Joss, \& Hsu 1992; Wellstein, Langer, \& Braun 2001; but see also Braun \& 
Langer 1995).  On the other hand, if accretion occurs after the 
secondary has already exhausted hydrogen in its core, accretion increases the mass of the 
envelope, but does not affect the mass of the core.  This changes the subsequent evolution of 
the secondary, since it is likely to spend the rest of its life as a blue rather than red 
supergiant \citep{podsi89}.  The latter accretion scenario has been proposed to explain the
blue supergiant progenitor of SN 1987A.  However, we do not consider this channel in the
present study, since it accounts for only a few percent of massive primordial binaries.  

Dynamically unstable mass transfer is accompanied by a common-envelope (CE) phase
and a spiral-in of the secondary through the envelope of the primary.  We use a 
standard, simple energy relation to determine the orbital separation following the 
spiral-in \citep[e.g.,][]{webbink84,dewi00}:
\begin{equation}\label{eq:ce}
\frac{a'}{a} = \frac{M_c M_2}{M_1} 
\left( M_2 + \frac{M_e}{2 \ace \, \lambda \, r_{\rm L1}} \right)^{-1} ~.
\end{equation}
The constants $\lambda$ and $\ace$ parameterize, respectively, the structure of the 
primary at the onset of Roche lobe overflow and the efficiency with which orbital binding 
energy is used to eject the CE.  For our standard model, we 
choose $\eta_{\rm CE} = 1.0$ and $\lambda = 0.5$ \citep[see][]{dewi00}.  If the secondary 
fills its Roche lobe for the computed final orbital separation, we assume that the binary 
components have merged.  For binaries that undergo unstable case B$_e$ and C$_e$ mass 
transfer, where $q < q_{\rm crit}$, we find that a merger occurs in nearly every instance.    

In all cases where a stellar merger is avoided, we assume that the entire 
hydrogen-rich envelope of the primary is removed.  By the time the primary
reaches the base of the first giant branch (beginning of case B$_l$ evolution) its
core is well developed, with a mass given approximately by \citep[e.g.,][]{hurley00}
\begin{equation}
M_c \simeq 0.1\,M_1^{1.35} ~,
\end{equation}
where $M_c$ and $M_1$ are in solar units.
We assume that this is the mass of the helium core immediately following case B$_e$ 
mass transfer as well, although it may be somewhat smaller, since mass transfer
interrupts the evolution of the primary \citep{wellstein01}.  For case C and D evolution, 
the mass of the core may be larger by $\sim 0.5 - 1 \msun$ as a result of shell nuclear 
burning.  If $M_c \la 3 \msun$ following case B mass transfer, the remaining helium star 
may grow to giant dimensions \citep{habets86b} upon central helium exhaustion and possibly 
fill its Roche lobe, initiating a phase of so-called case BB mass transfer 
\citep{degreve77,delgado81,habets86a}.

At the end of the mass-transfer phase, the result should be a stellar merger or a binary 
consisting of the secondary and the core of the primary.  Subsequently, the remaining 
nuclear fuel in the primary's core is consumed, leading to core collapse and a SN explosion.  
The post-SN orbital parameters are computed by taking into account the mass lost from the 
primary and the kick delivered to the newly-formed NS.  In our simulations, we neglect 
the effect of the SN blast wave on the secondary.  The mathematical formalism that we
utilize to compute the post-SN orbital parameters is outlined in Appendix B of PRP.

The further evolution of the system depends on the orbital separation and the mass of 
the secondary following the SN, as well as the degree to which the secondary has been 
rejuvenated after it has accreted mass.  If the secondary is of low or intermediate mass 
($\la 4 \msun$) and the periastron separation of the post-SN orbit is not too large,
the system will spend some time as a low- or intermediate-mass X-ray binary 
\citep[e.g.,][]{podsi01} after the orbit circularizes and the secondary grows to fill its 
Roche lobe.  If the secondary is massive, its strong stellar wind may allow the system to 
be detected as a HMXB.  However, the extreme mass ratio guarantees that a CE 
and spiral-in will occur not long after the secondary fills its Roche lobe.  

For this latter case where the secondary is massive, the final outcome of the spiral-in may be a merger, 
resulting in the formation of a Thorne-\.Zytkow (1975, 1977) object, or the successful dispersal 
of the common envelope (if the orbital period after the first SN is $\ga 100\day$;
e.g., Taam, Bodenheimer, \& Ostriker 1978).  If collapse to a black hole, via ``hypercritical''
accretion \citep[e.g.,][]{chevalier93,fryer96,brown00}, and merger are avoided, the NS will emerge
in a tight orbit with the hydrogen-exhausted core of the secondary.  A double NS binary
is then formed if the system remains bound following the supernova explosion of the secondary's core.  
In \S~\ref{sec:dns}, we discuss the implications of the new class of HMXBs for the formation
of double NSs.

%%%%%%%%%%%%%%%%%%%%%%%%%%%%%%%%%%%%%%%%%%%%%%%%%%%%%%%%%%%%%%%%%%%

\section{THE STATISTICAL SIGNIFICANCE OF THE \\ NEW CLASS OF BINARIES}\label{sec:sig}

Before we begin to explore alternative models to explain the new class of HMXBs
with wide orbits and small or moderate eccentricities, it is important that we provide 
some reasonable confirmation that this class is really a distinct population and does
not fit within the conventional framework of massive binary population synthesis.  It is, 
of course, possible that these systems are not dynamically significant (e.g., that their low 
eccentricities are result of tidal circularization), or that they represent the 
tail of a distribution and that some observational bias favors their detection.

\subsection{Tidal Circularization}

The high, persistent X-ray luminosities of the short-period HMXBs 
($P_{\rm orb} \la 10\day$) are probably maintained by a secondaries that are filling, 
or nearly filling, their Roche lobes.  This hypothesis is supported by the ellipsoidal 
variations of the optical lightcurves a number of these sources, which indicate that the stellar 
companions are tidally distorted.  Therefore, strong tidal interactions can easily explain the 
low eccentricities seen among the short-period HMXBs.  However, it is extremely unlikely 
that tidal circularization has played a significant role in modifying the orbits of the new 
class of wide, nearly circular HMXBs. 

Efficient tidal circularization requires that the star almost fills its Roche lobe 
and that there be an effective mechanism for damping the tide.  These conditions
are encapsulated by the cicularization timescale,  
\begin{equation}
\tau_{\rm cir} \propto \tau_{\rm dis} \left(\frac{a}{R}\right)^8 ~,
\end{equation}
in the limit of weak tidal friction \citep[e.g.,][]{zahn77,rieutord97}, where $a$ is the semimajor axis 
of the orbit, $R$ is the radius of the star, and $\tau_{\rm dis}$ is the timescale associated 
with viscous dissipation in the star.  Radiative dissipation of the tidal luminosity is 
enormously inefficient in comparison to turbulent dissipation in a convection zone.  
As a result, the tidal coupling to the radiative envelope of a massive main-sequence star 
is much weaker than the coupling to the convective core \citep{zahn77}.  However, since the core only 
comprises $\sim 20\%$ of the radius of the star, the resulting circularization timescale is comparable 
to, or shorter than, the star's nuclear lifetime only when the star is very nearly filling its Roche 
lobe.  This statement is supported by the near circularity ($e \la 3\times 10^{-3}$) of the orbits of 
SMC X-1, LMC X-4, and Cen X-3, which have periods less than 4\,days \citep{bildsten97,levine00}.  
However, tidal torques should have little effect on the orbit of a HMXB with 
$P_{\rm orb} \ga 10\day$, as long as the secondary is not too evolved and the eccentricity 
is not so large that the tidal interaction is enhanced dramatically at periastron.  

\begin{inlinefigure}
\centerline{\epsfig{file=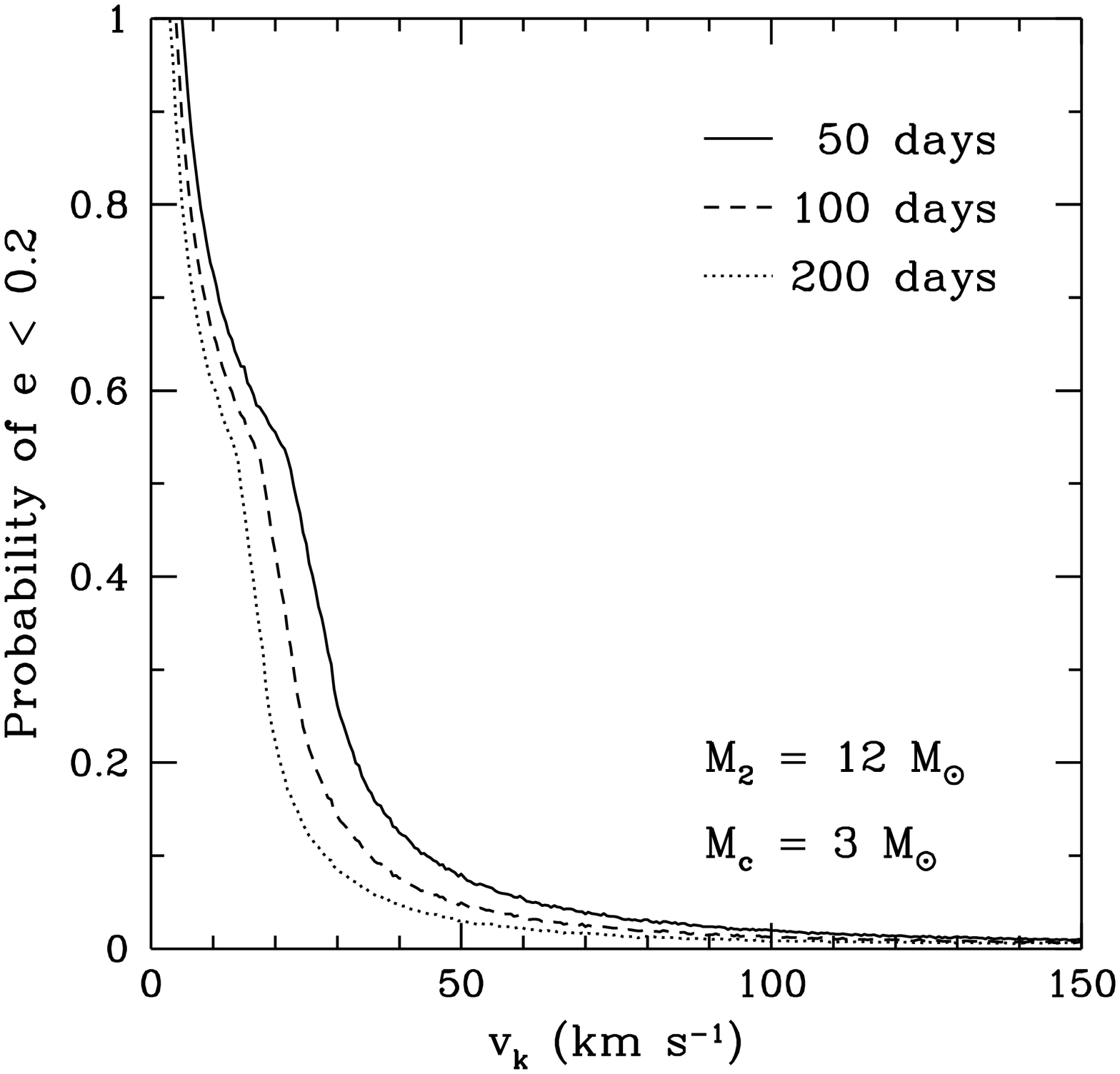,width=0.99\linewidth}}
\caption{Probability that the post-SN eccentricity is $<0.2$ for a helium star of 
mass $3\msun$ and a secondary of mass $12\msun$, for three different pre-SN orbital 
periods, as a function of the kick speed.  The directions of the kicks are distributed 
isotropically.}
\label{fig:pcir}
\end{inlinefigure}

\begin{figure*}[t]
\begin{minipage}[b]{0.47\linewidth}
\centerline{\epsfig{file=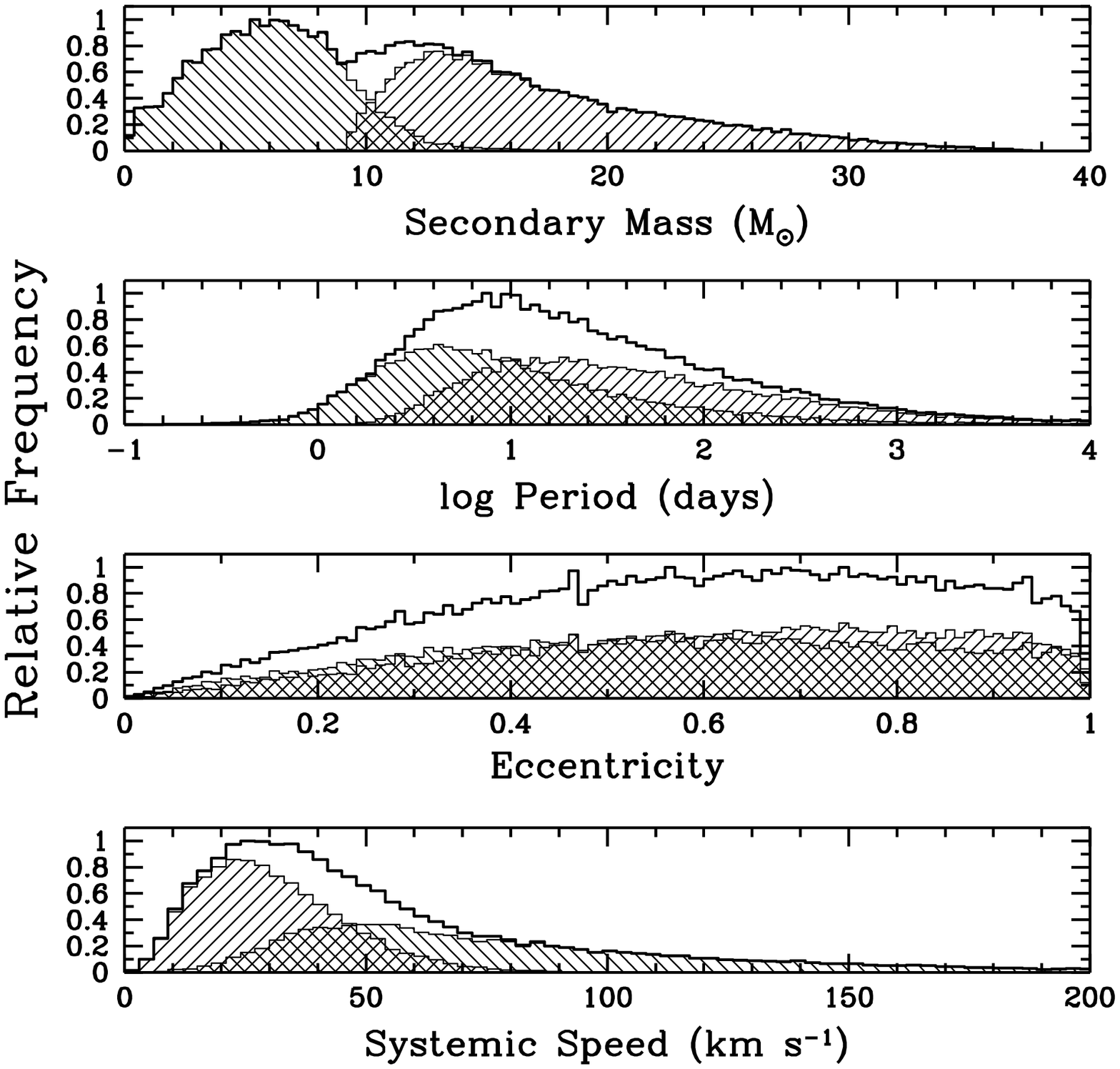,width=3.5in}}
\caption{Distributions of binary parameters of systems that have undergone case B or C mass transfer from 
the original primary star to the secondary, have been left bound following the supernova explosion, and have 
not merged to form a T\.ZO (see text).  Hatched regions indicate systems that have undergone stable mass 
transfer (+$45\degr$) and dynamically unstable mass transfer (--$45\degr$).  The histogram that encloses the 
hatched region is the sum of the distributions for stable and unstable systems.  A single Maxwellian kick 
distribution with $\sigma = 200\kms$ has been applied to all NSs. \vspace{15mm}}
\label{fig:200+200_hist}
\end{minipage}
\hfill
\begin{minipage}[b]{0.47\linewidth}
\centerline{\epsfig{file=hmxb_scat_200+200_col.ps,height=3.45in,angle=-90}}
\caption{Two-dimensional distribution of orbital period and eccentricity for the
systems in Fig. \ref{fig:200+200_hist} with secondary masses $>8\msun$.  The intensity and color
of a given square ``pixel'' indicate the total number of binaries and the stability of mass
transfer, respectively.  Pure red indicates only stable mass transfer, while pure blue indicates
only dynamically unstable mass transfer.  Colors that are not blue or red represent a mixture of
stable and unstable systems; pure green indicates equal numbers.  The intensity scale is linear
and the lightest pixels represent $\sim 1\%$ of the number represented by the most intense
pixel in the figure.  Overlayed on the plot are markers that show the period and eccentricity 
(or upper limit) for the wide, low-eccentricity HMXBs (circles), the better-known eccentric HMXBs 
(triangles), and the long-period binary radio pulsars with massive companions (squares).}  
\label{fig:200+200_scat}
\end{minipage}
\end{figure*}

\subsection{Observational Selection Effects}\label{sec:obs}

There are currently six candidates for the class of wide, low-eccentricity HMXBs (Table 1), 
four of which have identified O or B optical counterparts.  If we count all six candidates, 
then the new class of binaries accounts for roughly $30\%$ of the HMXBs with measured orbital 
parameters.  This fairly large fraction suggests that either these systems are preferentially 
selected for purely observational reasons or that their intrinsic population is actually quite large.

We should expect that an X-ray pulsar in a wide orbit with a low eccentricity is more difficult to 
detect and measure than if the orbital period is relatively short.  There are two primary reasons for 
this.  First, Bondi-Hoyle theory \citep{bondi44} predicts that the persistent luminosity of a wind-fed 
X-ray pulsar decreases with increasing orbital period, for a given rate of mass loss from the stellar 
companion.  Therefore, very wide binaries have a small effective Galactic volume in which their 
orbits are readily measurable; e.g., for low-luminosity sources that resemble X Per/4U 0352+309, 
it would currently be difficult to determine orbits if the systems lie much beyond 
1\,kpc.  Second, an accurate determination of the orbit from X-ray timing requires a series of 
observations that cover at least one full orbital cycle.  If the orbital period is very long, 
this may not be feasible, especially given the transient behavior of many sources and the limited
amounts of observing time.  Transient sources, of which there are four in Table 1, may be very 
conspicuous during outburst, but because of their transient nature and possibly large pulse-period 
variations (due to accretion torque noise), it can be difficult to measure 
their orbits very precisely.      

\subsection{The Case for Small Kicks: \\ Preliminary Arguments}

The most important factor in determining from model calculations the number of wide, 
low-eccentricity HMXBs in the Galaxy is the distribution in NS natal kick speeds.  A kick 
speed that is comparable to the relative orbital speed prior to the SN is likely to yield a 
highly eccentric binary following the explosion, including the probable event that the 
binary is disrupted ($e>1$).  This statement is illustrated more quantitatively in 
Fig.~\ref{fig:pcir}, where we plot the probability that the post-SN eccentricity is $< 0.2$ for a 
$3 \msun$ helium-star primary (NS progenitor) and a $12 \msun$ secondary (typical pre-SN masses) 
with one of three pre-SN orbital periods, as a function of the kick speed, $v_k$.  The distribution 
in kick directions was assumed to be isotropic.  For all three initial orbital periods, the 
probability is $< 5\%$ when $v_k > 0.5\,v_{\rm orb}$, and the probability is $<10\%$ when 
$v_k > 50 \kms$.  

\citet{hansen97} found that a Maxwellian distribution in kick speeds, given by 
\begin{equation}
p(v_k) = \sqrt{\frac{2}{\pi}} \frac{v_k^2}{\sigma^3} e^{-v_k^2/2\sigma^2}~, 
\end{equation}
is consistent with the data on pulsar proper motions, where the best fit
was obtained with $\sigma \simeq 190 \kms$.  A recent study by Arzoumanian, Chernoff, \& 
Cordes (2001) utilized a two-component Maxwellian distribution, and found that $40\%$ of their
model pulsars were contained in the low-speed component with $\sigma \sim 90\kms$, while the remaining
pulsars populated the high-speed component with $\sigma \sim 500\kms$. A single-component Maxwellian 
kick distribution predicts that $\sim 3\%$ of NSs are born with speeds $< 50 \kms$ for 
$\sigma = 100 \kms$, and $\sim 0.4\%$ for $\sigma = 200 \kms$.  This information, combined with the 
results displayed in Fig. 1, illustrates that the conventional wisdom regarding NS kicks does not 
favorably produce wide binaries with low eccentricities.  However, the statistical significance of 
the new class of HMXBs can only be quantitatively demonstrated by combining a complete population 
synthesis study with observational considerations regarding the discoverability of these systems.  

\subsection{The Case for Small Kicks: \\ Population Study}

Our population synthesis yields the fraction, $F_0$, of massive, primordial binaries that 
evolve into incipient HMXBs with orbital parameters in the range $P_{\rm orb} > 30\day$ and 
$e < 0.2$.  An upper limit to the expected present total number of HMXBs in the Galaxy with properties
similar to those in Table 1 is obtained if we multiply $F_0$ by the Galactic formation rate of 
massive stars, $\sim 10^{-2} \yr^{-1}$ \citep[the approximate Galactic rate of core-collapse 
supernovae;][]{cappellaro99}, and the maximum lifetime of the HMXB phase, $\sim 10^7\yr$ (the 
approximate evolutionary timescale of the secondary).  Therefore, the current total number of wide, 
low-eccentricity HMXBs in the Galaxy is expected to be $N_{\rm tot} < F_0 \times 10^5$.            

Of course, only a fraction, $F_{\rm dis}$, of the $N_{\rm tot}$ HMXBs could have been 
discovered by X-ray satellites that have scanned and/or monitored the X-ray sky (e.g., {\em Uhuru},
{\em HEAO-1}, {\em RXTE}, {\em CGRO}).  A simple way to estimate $F_{\rm dis}$ is to apply a flux limit,
$S_{\rm min}$, that is appropriate for a particular satellite instrument.  For a given X-ray 
luminosity, $\lx$, the maximum distance at which the source could be detected is 
$d_{\rm max} = (L_{\rm X}/4 \pi S_{\rm min})^{1/2}$.  If we assume that HMXBs do not move
very far from where they are formed, then an estimate of $F_{\rm dis}$ for a population of sources
of luminosity $\lx$ is just the probability that an O or B star is formed in a cylinder of 
radius $d_{\rm max}$ about the position of the Sun, perpendicular to the Galactic plane.  
Following \citet{paczynski90} and \citet{brandt95}, we adopt a disk distribution of stars 
given by
\begin{equation}\label{eq:form}
p(R) \propto R \exp (-R/R_0) ~,
\end{equation}
where $R$ is the Galactocentric radius and $R_0$ is the radial scalelength, taken to be
$4.5 \kpc$ \citep{kruit87}.  

The sensitivity with which the 2-10\,keV X-ray sky has been probed for weak and transient 
HMXB systems is difficult to estimate.  Some early scanning detectors aboard {\em Uhuru} 
and {\em HEAO-1} surveyed the sky for relatively brief periods (e.g., $\sim 1\yr$) with 
detection sensitivities as low as 
$S_{\rm min} \simeq 6 \times 10^{-11} {\rm ergs\,s^{-1}\,cm^{-2}}$.  In more recent times, 
the {\em Ginga} and {\em RXTE} satellites have been used to conduct limited pointed surveys 
of small regions of the sky, searching for X-ray pulsations from HMXBs; such studies were 
sensitive down to $S_{\rm min} \simeq 3 \times 10^{-11} {\rm ergs\,s^{-1}\,cm^{-2}}$.
However, the most sustained survey of the sky, with reasonable sensitivity, is that being 
conducted by the ASM aboard the {\em RXTE} satellite, which has been operating 
successfully for the past 5 years.  It has sensitivities of 
$S_{\rm min} \simeq 3 \times 10^{-10} {\rm ergs\,s^{-1}\,cm^{-2}}$ for X-ray sources of known 
position, and $S_{\rm min} \simeq 2 \times 10^{-9} {\rm ergs\,s^{-1}\,cm^{-2}}$ for the 
detection of new sources (e.g., transients).  

Sources of steady luminosity comparable to 
that of X Per and/or $\gamma$ Cas (i.e., $\sim 10^{35}\ergs$) would likely have 
been detected in previous surveys of the sky out to distances of $\sim 3\kpc$.  However, 
objects that are transient in nature, with only infrequent ``on'' states at these low 
luminosities, might be detected with the ASM only out to distances of $\sim 600\pc$.
Thus, the fractional effective volume of our Galaxy (from eq.~[\ref{eq:form}]) that has been 
well studied for wide, low-luminosity HMXBs probably lies in the range of 
$\sim 10^{-4} - 5\times10^{-3}$.  This is the range of values that we then consider for our parameter 
$F_{\rm dis}$.  Of course, if transient X-ray sources flare up to much higher luminosities, 
then the discovery probability at larger distances can go up dramatically.

Some results of our population study are shown in Figs.~\ref{fig:200+200_hist} 
and \ref{fig:200+200_scat}, where we have adopted a Maxwellian kick distribution
with $\sigma = 200 \kms$ and the standard-model parameters given in \S~\ref{sec:popsyn}.  
It is apparent from Figs.~\ref{fig:200+200_hist} and \ref{fig:200+200_scat} that binaries 
containing a massive secondary and that have low eccentricities and long periods are not 
produced favorably.  These simulations yield $F_0 \sim 4\times 10^{-4}$, which corresponds to at 
most $\sim 40$ wide, nearly circular HMXBs in the Galaxy.  It is not at all reasonable to suspect 
that we have already detected and precisely measured the orbits of $\sim 15\%$ (i.e., 6 of 40)
of this inconspicuous population, over the entire Galaxy.  For a reduced value of 
$\sigma = 100 \kms$, we find that $F_0$ is increased by roughly a factor of five, and so perhaps 
as many as 200 such objects are present in the Galaxy.  This again is probably too small a number, 
considering that the assumed detectable lifetime of $10^7\yr$ is likely to be an upper limit, and 
that the efficiency for discovery, $F_{\rm dis}$, must be quite low.  

A Maxwellian kick distribution with $\sigma \sim 100 \kms$, applied uniformly to {\em all} NSs, 
may be consistent with the speeds of a large fraction of isolated pulsars with measured proper
motions \citep{arzoumanian01}, but conflicts arise when we consider significantly smaller 
values.  The kinematics of the populations of single pulsars and LMXBs do suggest a large mean 
kick speed \citep[e.g.,][]{hansen97,brandt95,johnston96}, and a Maxwellian distribution with 
$\sigma \ga 100 \kms$ seems to reproduce the properties of these populations reasonably well.  
However, based upon our discussion above, we suggest that there are many wide, nearly circular 
HMXBs in the Galaxy (possibly well in excess of 1000), and that the NSs in these systems require 
fairly small kicks ($v_k \la 50 \kms$) on average\footnote{Note that the mean of a Maxwellian
distribution is given by $(8/\pi)^{1/2} \sigma$.}.  The apparent conflict with the other known
NS populations is resolved if the mean kick speed depends on the evolutionary history of the NS 
progenitor in a binary system.  We now go on to describe this scenario in the next two sections.    

%%%%%%%%%%%%%%%%%%%%%%%%%%%%%%%%%%%%%%%%%%%%%%%%%%%%%%%%%%%%%%%%%%%

\section{AN EVOLUTIONARY MODEL}\label{sec:mod}

With some perspective, we can motivate a phenomenological picture that accounts
for the new population of long-period, low-eccentricity HMXBs, and which is consistent 
with what we know about the Galactic NS populations.  There are four basic constraints 
that our model must satisfy.  First, the orbits of the systems listed 
in Table 1 suggest that they did not experience a dynamical spiral-in phase prior to 
the first SN, and that the NSs in these binaries did not receive a very large kick.
We propose that a significant fraction of those NSs whose progenitors underwent 
case B$_e$ or C$_e$ mass transfer received natal kick speeds of 
$\la 50 \kms$.  Second, the orbits of all other binaries containing a NS and a massive 
stellar companion should be naturally accounted for.  Such binaries include 
short-period HMXBs and moderately wide, eccentric HMXBs (see \S~\ref{sec:newbin}), as well
as the three long-period binary radio pulsars with massive companions 
\citep[PSR B1259--63, PSR J1740--3052, and J0045--7319;][]{johnston92,kaspi94,manchester95,kaspi96,stairs01}.
Third, the model should be able to approximately reproduce the numbers and properties of luminous low-mass 
X-ray binaries in the Galaxy.  Fourth, our basic picture should also be consistent with the 
observed kinematical distribution of isolated pulsars in the Galaxy, on which the NS kick
distributions are based.  

The orbits of the observed short-period HMXBs have been affected by tidal interactions
(see \S~\ref{sec:sig}), and so tell us very little about the NS kick.  HMXBs with orbital 
parameters of $P_{\rm orb} \sim 20 - 100 \day$ and $e \sim 0.3 - 0.5$, in addition to the long-period, 
binary radio pulsars are somewhat difficult to interpret definitively, since they are consistent a priori 
with being the products of either stable or dynamically unstable mass transfer.  This is indicated in 
Fig.~2 by the overlap of the period distributions for the stable and unstable systems.
If these binaries have experienced a dynamical spiral-in, then their survival (as opposed to merger) 
essentially requires that the mass transfer was case B$_l$ or C$_l$ (see \S~\ref{sec:popsyn}), and our 
model predicts that the NSs received the conventional large kicks.  If the mass transfer was stable 
(case B$_e$ or C$_e$), then a significant eccentricity is still possible as long as the magnitude of the 
kick is an appreciable fraction of the pre-SN orbital speed.  This point is important, and it is worthwhile
to discuss a particular example.  

Consider the very long-period, highly eccentric binary pulsar PSR B1259--63 with 
$P_{\rm orb} = 1236.72\day$ and $e = 0.87$.  The orbital separation at periastron is 
$\sim 140\rsun$, for an assumed mass of $10\msun$ for the secondary.  This is the smallest circular pre-SN 
orbit that is permitted \citep[e.g.,][]{flannery75}, and so the largest pre-SN relative orbital 
speed is $v_{\rm orb} \sim 130-170\kms$, for a reasonable range in pre-collapse core masses.  
If the fractional mass lost in the SN explosion is small, then a post-SN eccentricity of order 
unity is possible for a kick speed that is $\la 40\%$ of the orbital speed, or $\la 70\kms$ for 
PSR B1259--63 (e.g., Brandt \& Podsiadlowki 1995; Appendix A of PRP); in an absolute sense, this 
is not a very large kick.  

The latter two semi-empirical constraints on our model are satisfied if we suppose that a 
NS receives the usual large kick if its progenitor is allowed to evolve into a red supergiant 
(i.e., a single progenitor or case B$_l$, C$_l$, or D for a binary system).  Within this 
framework, isolated, fast-moving pulsars are likely to have come from single progenitors or 
wide binaries that were disrupted by the SN explosion.  Also, by our hypothesis, the NSs born in 
LMXBs would receive kicks drawn from a conventional distribution, since their standard formation 
channel involves a common-envelope phase in the case B$_l$ or C$_l$ scenario 
\citep[e.g.,][]{bhat91,kalogera98a}.               

We have redone our population synthesis calculation with the following simple modification.
If the mass transfer is case B$_e$ or C$_e$ (see \S~\ref{sec:popsyn}), the NS kick is 
chosen from a Maxwellian distribution with $\sigma = 20 \kms$, a small but otherwise
arbitrary value.  On the other hand, if the mass transfer begins while the primary is a red 
supergiant (case B$_l$, C$_l$), or there is no mass transfer (case D), we adopt a more 
conventional kick distribution, with $\sigma = 200 \kms$.  Rather than applying the same 
kick distribution to {\em all} case B$_e$ and C$_e$ systems, we could have focused our 
attention on only the stable case B$_e$ and C$_e$ systems, since these are the alleged 
progenitors of the new class of HMXBs.  However, our choice of treating all case B$_e$ and 
C$_e$ binaries on an equal footing is partly motivated by a theoretical model, which we discuss 
in the next section.   

\begin{figure*}
\begin{minipage}[b]{0.47\linewidth}
\centerline{\epsfig{file=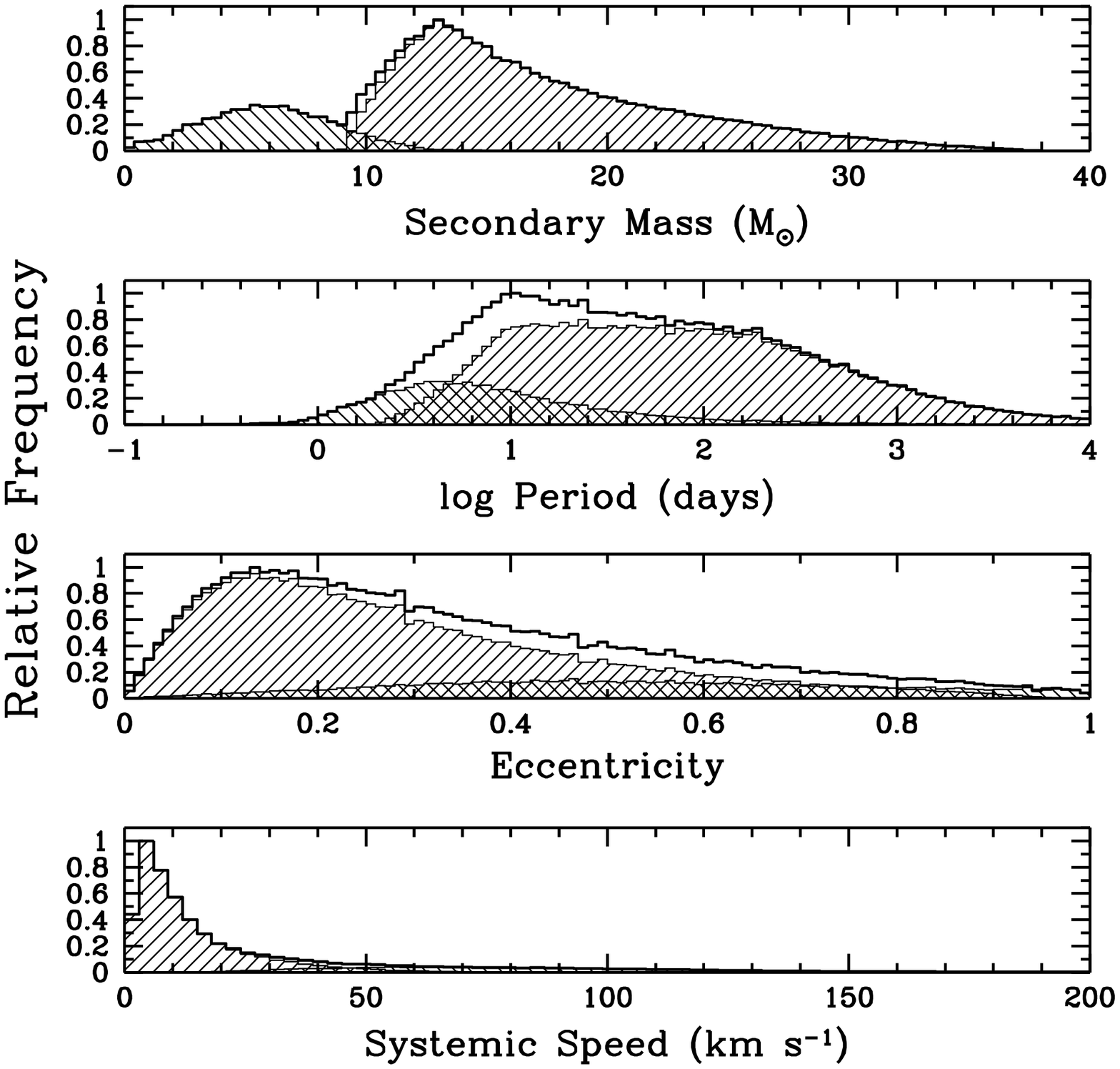,width=3.5in}}
\caption{Same as Fig. \ref{fig:200+200_hist}, but with $\sigma = 20\kms$ applied to the NSs born
in all case B$_e$ and C$_e$ binaries, and $\sigma = 200$ applied to all other NSs.
Note that the eccentricity distribution for the stable systems has a distinct peak at 
$e \sim 0.15$.}
\label{fig:200+20_hist}
\end{minipage}
\hfill
\begin{minipage}[b]{0.47\linewidth}
\centerline{\epsfig{file=hmxb_scat_200+20_col.ps,height=3.5in,angle=-90}}
\caption{Distribution of orbital period and eccentricity for the systems in Fig. \ref{fig:200+20_hist} 
with secondary masses $>8\msun$.  A value of $\sigma = 20\kms$ applied to the NSs born
in all case B$_e$ and C$_e$ binaries, and $\sigma = 200$ applied to all other NSs.  The colors, intensities,
and symbols have the same meaning as in \ref{fig:200+200_scat}.}
\label{fig:200+20_scat}
\end{minipage}
\end{figure*}

Figures~\ref{fig:200+20_hist} and \ref{fig:200+20_scat} should be compared to 
Figs.~\ref{fig:200+200_hist} and \ref{fig:200+200_scat}, respectively.  Many more systems with 
long periods and low eccentricities are produced when $\sigma = 20 \kms$ is adopted for 
the case B$_e$ and C$_e$ systems.  This simulation yields $F_0 \sim 2.6\times 10^{-2}$ and 
a present Galactic population of $\la 2600$ wide, low-eccentricity HMXBs, a factor of 65 
more than in the case where $\sigma = 200 \kms$ is applied to all NSs.  If such a large 
population of long-period HMXBs is indeed present in the Galaxy, their integrated X-ray 
luminosity cannot exceed $\sim 10^{38}\ergs$, the total X-ray luminosity of the so-called 
``Galactic ridge'' of unresolved X-ray sources \citep[e.g.,][]{yamasaki97,valinia98,valinia00}.  
This is not problematic if these HMXBs typically have persistent X-ray luminosities of 
$\la 10^{35} \ergs$.

We have also calculated the production efficiency for systems that may evolve to 
resemble the HMXBs with moderate-to-long periods and significant eccentricities 
(triangles in Figs.~\ref{fig:200+200_scat} and \ref{fig:200+20_scat}), as well 
as for systems similar to the massive, long-period, highly eccentric, binary radio pulsars  
(squares in Figs.~\ref{fig:200+200_scat} and \ref{fig:200+20_scat}).  In our code, we 
simply defined the HMXBs by the parameter ranges of $20\day < P_{\rm orb} < 100\day$ and 
$0.3 < e < 0.5$, and the binary radio pulsars by $100\day < P_{\rm orb} < 1000\day$ and $0.5 < e < 0.9$.  
Furthermore, we define the formation efficiency as the fraction of primordial binaries 
that ultimately evolve into the systems of interest (the parameter $F_0$ for the new class of HMXBs).  
If we apply the conventional kick scenario, with $\sigma = 200\kms$ for all NSs, the formation efficiencies are 
$\sim 0.4\%$ for both the eccentric HMXBs and binary radio pulsars.  On ther other hand, in our modified 
kick scenario described above, the formation efficiency for the eccentric HMXBs is $\sim 1.7\%$, 
while for the binary radio pulsars the efficiency is $\sim 1.3\%$.  The increase in the number of systems 
is certainly substantial, but not nearly as dramatic as the increase in the number of long-period, 
low-eccentricity HMXBs.  

%%%%%%%%%%%%%%%%%%%%%%%%%%%%%%%%%%%%%%%%%%%%%%%%%%%%%%%%%%%%%%%%%%%

\section{A PHYSICAL MODEL}\label{sec:phys}

The simple scenario we have outlined above is purely phenomenological.  If the 
picture is essentially correct, then we should ask: What physical process(es) 
may account for the dependence of the NS kick on the evolutionary history 
of its progenitor in a binary system?  We suggest that the rotation of the 
collapsing core plays a crucial role in determining the magnitude of the NS
kick, and that there is a natural reason to expect a possibly sharp break in the 
distribution of rotation rates of stellar cores exposed following mass 
transfer.  

Many young, isolated, massive stars are observed to rotate at $\sim 20-50 \%$ of their 
breakup rates \citep[e.g.,][]{fukuda82,howarth97}.  For a main-sequence star of mass 
$10\msun$, the breakup angular frequency is $\Omega_b \sim 10^{-4}\,{\rm rad\,s}^{-1}$.
If the stellar core initially has the same angular velocity, and the core retains a constant
angular momentum as it evolves, then the NS that is produced is expected to rotate close to its
breakup rate (i.e., with a period of $\la 1\,{\rm ms}$).  However, the question of exactly how such 
rapid rotation on the main sequence translates to the rotation of the pre-collapse iron core, 
immediately prior to NS formation, is difficult to answer, owing to the large number and 
complexity of hydrodynamical and magnetohydrodynamical angular-momentum transport processes.  
Heger, Langer, \& Woosley (2000; hereafter, HLW) have conducted the most sophisticated and 
detailed study to-date of isolated, rotating, massive stars, which included treatments of various 
hydrodynamical instabilities, but neglected the influence of magnetic fields.  Interestingly, 
they found that the angular momentum of the pre-collapse core was quite insensitive to the initial 
rotation rate of the star, and that the nascent NS remnant would spin at close to its breakup rate, 
although their simulations could not follow the evolution beyond the start of core collapse.    

In the binary systems we are considering, the first phase of mass transfer is expected to 
strip the hydrogen-rich envelope from the primary (see \S~\ref{sec:popsyn}), thus exposing 
its core.  The results of HLW suggest that this core will be a fairly rapid 
rotator if the primary was initially rapidly rotating.  Furthermore, the results of HLW 
indicate that the angular momentum of the exposed core should not depend strongly on
the evolutionary state of the primary at the onset of mass transfer, if only hydrodynamical 
angular-momentum transport processes are considered.  However, magnetic fields may introduce 
just such a dependence.  

In the presence of differential rotation, an initially poloidal magnetic field will be 
wound up into a predominantly azimuthal field, where the magnitude of the azimuthal component 
is proportional to the number of differential turns (for a review of this and related 
magnetohydrodynamic processes in stars, see Spruit 1999).  If this 
generation and amplification of the magnetic field occurs between the surface of the convective 
core and the outlying stellar envelope of a massive star, then the resulting magnetic torque will 
cause the core to spin down.  The torque is transmitted by the $r$-$\phi$ component, 
$B_r B_\phi / 4 \pi$, of the Maxwell stress tensor.  Suppose that the core of mass $M_c$ is 
initially rotating with an angular velocity, $\Omega_c$, and that the stellar envelope 
is nonrotating.  The timescale for the core to spin down, $\tau_s$, is approximately 
\citep[see][]{spruit98b}
\begin{multline}\label{eq:mag}
\tau_s \sim \frac{I_c \Omega_c}{r_c^3 \bbar^2} 
\sim  10 \gyr 
\left(\frac{k}{0.1}\right) 
\left(\frac{M_c}{M_\odot}\right) \\
\times \left(\frac{\Omega_c}{10^{-4}\,{\rm s}^{-1}}\right) 
\left(\frac{\bbar}{1\,{\rm G}}\right)^{-2}~,
\end{multline}
where $I_c=k M_c R_c^2$ is the moment of inertia of the core and $\bbar = (B_r B_\phi)^{1/2}$.
The geometric mean field $\bbar$ appearing in eq.~(\ref{eq:mag}) increases with time, because of
the winding-up of the $B_r$ component.  The 1\,G field used for scaling in eq.~(\ref{eq:mag}) is 
quite small as compared with field strength that can, in principle, be reached as a result of the
amplification process.  It is thus possible for the coupling timescale $\tau_s$ to be
shorter than the evolutionary timescale of the star.

Immediately following the depletion of hydrogen in the core, a massive star expands to giant 
dimensions on a thermal timescale ($\sim 10^4-10^5 \yr$).  Significant differential rotation
between the core and the envelope will only be established after the star crosses the 
Hertzsprung gap and develops a deep convective envelope.  It is thus reasonable
to suggest that magnetic torques should not be very effective in spinning down the core during
this short-lived evolutionary phase, and that a helium star should be rapidly rotating if it is 
uncovered following case B$_e$ or C$_e$ mass transfer.       

If mass transfer takes place at a later stage of evolution (i.e., during the first giant branch or asymptotic
giant branch), the stellar core may rotate millions of times with respect to the very slowly 
rotating convective envelope.  One might take the view \citep{spruit98a} that under these circumstances 
there is plenty of time for a strong toroidal magnetic field to build up. The consequence of this would be 
that the cores of evolved stars would approach corotation with their envelopes, and their angular momentum 
would then be so small that the NSs formed will have spin periods of hundreds of seconds. This is problematic, 
since the spin of observed young NS are tens of milliseconds in several cases (e.g., the Crab pulsar has a period of 
33 ms).  \citet{spruit98a} resolved this dilemma by attributing the current, short spin periods to off-center kicks.

In reality, the magnitude of the geometric mean field $\bbar$ that can actually be obtained
in a differentially rotating star is not just a matter of simple winding-up of field lines. 
The approximately azimuthal fields that develop from differential rotation are known to be prone to 
instabilities \citep{tayler73,acheson78}.  These instabilities may limit the attainable field
strengths (for a discussion, see Spruit 1999).  On the other hand, the unstable fluid
displacements would create new poloidal field components which in turn would be wound
up to generate more azimuthal field. It is thus possible that an unstable azimuthal field will develop 
into a {\em dynamo process} operating on the differential rotation. In \citet{spruit01} an estimate 
is developed for the behavior of such a dynamo process and the $\bbar$ it produces. Preliminary 
calculations of the evolution of rotating stars that incorporate this formalism 
(Heger, Woosley, \& Spruit, in preparation) indicate that the coupling between cores and envelopes 
could be less efficient than assumed in \citet{spruit98a}.     

Based upon the physical arguments presented above, and the phenomenological picture discussed 
in \S~\ref{sec:mod}, we suggest that rapidly rotating stellar cores exposed following stable 
case ${\rm B}_e$ or ${\rm C}_e$ mass transfer produce NSs with small natal kicks, 
while NSs formed at a later stage of evolution (case B$_l$, C$_l$, or D), where the pre-collapse 
cores may be spinning quite slowly, receive the conventional large kicks.  The collapse of a rapidly 
rotating core is certainly more dynamically complex than the collapse of a core that is initially 
nonrotating.  However, it is not obvious a priori whether a rapidly or slowly spinning pre-collapse core 
should ultimately yield a larger average natal kick to the NS, since the physical mechanisms that may be
responsible for the kick are poorly understood.  One possibility is that rapid rotation has
the effect of averaging out the asymmetries that give rise to large NS kicks \citep{spruit98a}.

%%%%%%%%%%%%%%%%%%%%%%%%%%%%%%%%%%%%%%%%%%%%%%%%%%%%%%%%%%%%%%%%%%%

\section{FURTHER IMPLICATIONS OF THE MODEL}\label{sec:imp}

\subsection{Neutron Star Retention in Globular Clusters}

It is apparent that globular clusters must contain appreciable numbers of NSs.  
For example, 22 millisecond radio pulsars have been detected in the massive globular 
cluster 47 Tuc, and many more are thought to be present (Camilo et al. 2000; see also PRP 
and references therein).  This abundance of NSs raises an interesting question.  If NSs are 
born with speeds that are typically in excess of $100-200 \kms$, how is it that even a very 
dense globular cluster, with a central escape speed of $\sim 50\kms$, can retain so many?  
A conventional Maxwellian kick distribution, with $\sigma = 200 \kms$ applied to all 
NSs, predicts that only $\sim 0.4\%$ of NSs are born with speeds $< 50 \kms$, and $\sim 3\%$
with speeds $< 100 \kms$.  In PRP, we considered the influence of massive binary systems on 
the NS retention fraction.  Our standard model calculation showed that $\la 5\%$ of NSs born
in binary systems could be retained in a typical cluster. 

This long-standing {\em retention problem} is clearly alleviated if there exists a 
population of NSs that are born with kick speeds $\la 50 \kms$, which is seemingly at 
odds with the large speeds inferred for isolated pulsars in the Galactic disk.
The scenario that we have proposed in \S~\ref{sec:mod} to account for the long orbital
periods and low eccentricities of the HMXBs listed in Table 1 is not in conflict with 
the speeds of the isolated pulsars, by construction.  Our hypothesis is that low-kick
NSs are preferentially born in certain binary systems, and thus these NSs are much more likely 
to remain bound to their companions following the SN.  If the secondary is massive, possibly 
as a result of accretion, the effect of the impulsive kick on a bound post-SN binary is
diluted considerably, thereby allowing the binary to be retained in the cluster.  Our 
simulations indicate that the NS retention fraction may be increased by more than a 
factor of four (to $\ga 20\%$) if we adopt the phenomenological picture outlined in 
\S~\ref{sec:mod} (see PRP for further details).    

\subsection{Formation of Double Neutron Star Binaries}\label{sec:dns}

A double NS (DNS) --- a binary comprised of two NSs --- seems like an improbable
object; however, five proposed DNSs have been detected in the Galaxy.  In all cases, only 
one of the components of the DNS is detected as a radio pulsar, and the other component is 
inferred to be a NS based on the mass function.  The DNS in the globular cluster
M15, PSR 2127+11C, probably formed dynamically \citep[e.g.,][]{phinney91}, rather than from a 
massive primordial binary.  The present discussion is restricted to the formation 
of DNSs in the Galactic disk, where the dynamical formation of binaries does not occur
with any significant probability.

At the end of \S~\ref{sec:popsyn}, we very briefly described the standard formation 
scenario for DNSs in the Galactic disk (see, e.g., Bhattacharya \& van den Heuvel 1991 
for a more detailed discussion).  We stated that the envelope of the secondary can only
be successfully ejected by the first-formed NS if the orbit is sufficiently wide 
($P_{\rm orb} \ga 100\day$) at the time the secondary fills its Roche lobe.  Therefore, 
the episode of mass transfer before the first SN must have been stable for the 
majority of binaries that ultimately evolve into DNSs.  Our phenomenological picture
for the formation of wide, low-eccentricity HMXBs involves relatively low kick speeds applied 
to NSs born in binary systems that have undergone case B$_e$ or C$_e$ mass transfer.  For about 
half of the case B$_e$ and C$_e$ binaries (for $q_{\rm crit} = 0.5$; see \S~\ref{sec:popsyn}) 
the mass transfer is stable.  Thus, our model may result in a dramatically increased formation 
efficiency for DNS progenitors, since many more wide binaries remain bound following the first 
SN than if the conventional large kicks are applied to all NSs.   

We investigated the formation of DNSs with the following straightforward extensions to our
population synthesis code.  If the binary survives the first episode of mass transfer and
the first SN without merging and without being disrupted, then we consider the eccentric post-SN 
orbit of the first-formed NS and the secondary.  We suppose that once the secondary evolves to fill
its Roche lobe, the orbit quickly circularizes.  Because of the extreme mass ratio, the subsequent
phase of mass transfer is guaranteed to be dynamically unstable, and the orbital separation following
the spiral-in is computed using eq.~(\ref{eq:ce}).  If the new separation indicates that the 
radius of the hydrogen-exhausted core of the secondary exceeds its Roche lobe radius, then we assume
a coalescence is the result.  Finally, the new orbital parameters are computed following the SN
explosion of the secondary's core, where the kick to the second-formed NS is drawn from a 
Maxwellian distribution.  

It is interesting to note that, for the preferred progenitors of DNSs, 
the first episode of mass transfer was stable, in which case the secondary has accreted a considerable
amount of mass and angular momentum.  As a result, these secondaries should be rotating rapidly 
following mass transfer.  This seems to be borne out by observations of HMXBs, where many systems 
contain a Be optical counterpart; the Be phenomenon is likely a consequence of rapid rotation 
\citep[e.g.,][]{slettebak88}.
Therefore, it seems as though we may apply a kick to the second NS in precisely the same way as for 
the first, with a value of $\sigma$ that depends on the evolutionary state of the secondary when it 
fills its Roche lobe.  This turns out {\em not} to be very important, however, since the mass loss from the 
exploding core of the secondary typically has a more disruptive influence on the orbit than the kick. 

As a point of reference, we applied the more-or-less standard Maxwellian kick distribution,
with $\sigma = 200\kms$, to all NSs, both first- and second-formed.  We find that the 
fraction of primordial binaries that successfully evolve into DNSs is $\sim 10^{-3}$.  
For a core-collapse SN rate of $10^{-2}\yr^{-1}$, this fraction corresponds to an approximate DNS 
birthrate of $\sim 10^{-5}\yr^{-1}$, consistent with other recent theoretical calculations that 
used similar methods and assumptions \citep[e.g.,][]{lipunov97,zwart98}.  

If we assume that NSs born following case B$_e$ or C$_e$ mass transfer receive kicks drawn
from a Maxwellian with $\sigma = 20\kms$, we find that the DNS birthrate is increased by
roughly a factor of twenty, to $\sim 2 \times 10^{-4}\yr^{-1}$.  This order-of-magnitude 
increase is almost entirely accounted for by the increase in the number of viable DNS 
progenitors --- systems where the common-envelope is successfully ejected during the dynamical
mass transfer episode from the secondary to the first-formed NS.  

%%%%%%%%%%%%%%%%%%%%%%%%%%%%%%%%%%%%%%%%%%%%%%%%%%%%%%%%%%%%%%%%%%%

\section{SUMMARY}\label{sec:con}

Using a combination of observational and theoretical arguments, we have considered
the significance of a new observed class of HMXBs, with orbits that are distinguished 
by relatively long periods ($P_{\rm orb} \sim 30-250\day$) and low eccentricities
($e \la 0.2$).  Our analysis indicates that the conventional wisdom regarding NS
kicks does not adequately account for the number of these systems known at present,
which comprise roughly $30\%$ of HMXBs with measured orbital parameters.  Members of this new
class of HMXBs contain NSs that almost certainly received a fairly small kick
($\la 50\kms$) at the time of formation.  A prevalence of such low-kick NSs is simply
incompatible with large mean natal kick speeds ($\ga 200-300\kms$) inferred for isolated 
radio pulsars in the Galaxy.  However, we have developed a phenomenological model that 
simultaneously accounts for the long-period, low-eccentricity HMXBs and which does 
not violate any previous notions regarding the kinematics of other NS populations
(i.e., radio pulsars, LMXBs, and other HMXBs).  

Specifically, we propose that a NS receives a relatively small kick if its progenitor star
experienced case B$_e$ or C$_e$ mass transfer in a binary system.  In operational terms,
we utilized a Maxwellian distribution in kick speeds, but with a somewhat arbitrarily selected
low value of $\sigma = 20\kms$ applied to NSs born in case B$_e$ or C$_e$ binaries, and for all 
other NSs (case B$_l$, C$_l$, or D binaries, as well as isolated progenitors) we adopted a much 
higher value of $\sigma = 200\kms$.  This scenario results in sufficient numbers for the new
class of HMXBs, and, by construction, is consistent with the numbers and properties of other 
NS populations in the Galaxy.

If this phenomenological picture is basically correct, then there must be some physical
explanation for why the magnitude of the kick depends on the evolutionary history of the 
NS progenitor.  We suggest that the rotation of the pre-collapse core of a massive star 
introduces just such a dependence.  If the hydrogen-exhausted core of an initially rapidly 
rotating massive star is exposed following case B$_e$ or C$_e$ mass transfer in a binary,
then the core is also likely to be a rapid rotator, as implied by the work of 
\citet{heger00}.  On the other hand, if the NS progenitor is allowed to evolve into a 
red supergiant (case B$_l$, C$_l$, D, or a single star), then significant magnetic torques,
amplified by the strong differential rotation between the core and the deep convective 
envelope \citep{spruit98a,spruit99}, may cause the core to spin down dramatically.  Thus, 
for whatever reason, the dynamics of core collapse may be such that low kick speeds result 
for rapidly rotating pre-collapse cores, and cores that are spinning slowly preferentially
yield the conventional large kick speeds.
  
Our model to explain the new class of HMXBs requires that a large fraction of NSs born
in binary systems receive only a small recoil speed following core collapse and the
SN explosion.  This simple requirement has important implications for at least two
very different problems.  First, the problem of retaining NSs in globular clusters
is alleviated if not solved if our hypothesis is correct.  Second, our scenario predicts
an order of magnitude larger birthrate of double NS binaries than if the conventional
kick distributions are applied.  

%%%%%%%%%%%%%%%%%%%%%%%%%%%%%%%%%%%%%%%%%%%%%%%%%%%%%%%%%%%%%%%%%%%

\acknowledgements

This paper has benefitted from conversations with Simon Portegies Zwart, Vicky Kalogera, 
and Chris Fryer.  In addition, we would like to thank Duncan Galloway, Ed Morgan, and Al Levine 
for useful discussions and for providing us with some results of their recent work prior to publication.
EP would also like to acknowledge the hospitality of the Aspen Center for Physics, where some of this 
work was initiated.  This research was supported in part by NASA ATP grant NAG5-8368.  

%%%%%%%%%%%%%%%%%%%%%%%%%%%%%%%%%%%%%%%%%%%%%%%%%%%%%%%%%%%%%%%%%%%

%%%%%%%%%%%%%%%%%%%%%%%%%%%%%%%%%%%%%%%%%%%%%%%%%%%%%%%%%%%%%%%%%%%

\end{document}